\documentstyle[aps]{revtex}
\begin{document}
%\draft  % Makes pacs numbers print (REVTEX)
%
\title{Valence-bond states in dynamical Jahn-Teller molecular systems}
\vspace{10mm}
\author{Giuseppe Santoro$^{1,2}$, Leonardo Guidoni$^{1,2}$, 
Alberto Parola$^{1,3}$, and Erio Tosatti$^{1,2,4}$}
\address{$^{(1)}$ Istituto Nazionale per la Fisica della Materia (INFM), 
Via Beirut 2, Trieste, Italy\\
$^{(2)}$ International School for Advanced Studies (SISSA), Via Beirut 2,
Trieste, Italy\\
$^{(3)}$ Istituto di Scienze Fisiche, 
Universit\'a di Milano, via Lucini 3, Como, Italy\\
$^{(4)}$ International Center for Theoretical Physics (ICTP), Strada Costiera,
Trieste, Italy}
\date{\today}
\maketitle
\vspace{10mm}
\begin{abstract}
We discuss a hopping model of electrons between idealized molecular sites
with local orbital degeneracy and dynamical Jahn-Teller effect, for crystal 
field environments of sufficiently high symmetry.
For the Mott-insulating case (one electron per site and large Coulomb 
repulsions), in the simplest two-fold degenerate situation, we are led to 
consider a particular exchange hamiltonian, describing two isotropic spin-1/2 
Heisenberg problems coupled by a quartic term on equivalent bonds.
This twin-exchange hamiltonian applies to a physical regime in which the 
inter-orbital singlet is the lowest-energy intermediate state available for
hopping. 
This regime is favored by a relatively strong electron-phonon coupling.
Using variational arguments, a large-$n$ limit, and exact diagonalization
data, we find that the ground state, in the one dimensional case, 
is a solid valence bond state. 
The situation in the two dimensional case is less clear.
Finally, the behavior of the system upon hole doping is studied
in one dimension.
\end{abstract}
%
%\pacs{PACS Numbers: } % (REVTEX)
%
%%%%%%%%%%%%%%%%%%%%%%%%%%%%%%%%%%%%%%%%%%%%%%%%%%%%%%%%%%%%%%%%%%%%%%%%%%
%                               TEXT
%%%%%%%%%%%%%%%%%%%%%%%%%%%%%%%%%%%%%%%%%%%%%%%%%%%%%%%%%%%%%%%%%%%%%%%%%%
\newpage
\section{Introduction}

Ions or molecules with degenerate electronic orbitals coupled to local 
vibrations can display a rich variety of phenomena.\cite{KK} 
The most notable of these phenomena is the so-called static Jahn-Teller (JT) 
effect, i.e., the system gains energy by lowering the symmetry
through a static distortion, thereby removing the electronic degeneracy. 
For finite systems, like a molecule, quantum mechanics can play an important
role in that, if the frequency of the relevant vibronic modes is high enough,
it can dynamically restore the original symmetry through a quantum 
superposition of ``degenerate'' distorted configurations (dynamical JT effect).

Many systems containing JT ions or molecules are found to be insulators with
quite a rich variety of spin and orbital ordering effects. 
Such a richness of exchange effects is due to the interplay between the 
spin and orbital degrees of freedom.\cite{KK}

The simplest cases can be modeled by the so-called $E\otimes e$ JT-coupling, 
i.e., at each site $r$ a two-fold degenerate electronic orbital is coupled to a 
two-fold degenerate local vibron.\cite{Englman}
Although conceived as the simplest workable model, such a situation is not 
totally unrealistic, as higher molecular degeneracies can often 
be reduced, by the crystal field of the solid state environment, to 
residual doublets. 
(If the crystal field symmetry is very low, the degeneracy will be completely 
removed).

If exactly one electron sits on each JT site, and if the Coulomb repulsion
is very large, the system will have an insulating gap, due to Mott-Hubbard 
correlations, and the low-energy spectrum will consist of spin and orbital 
excitations. 
For the $E\otimes e$ case, which we are going to consider, the physics of 
the insulating state is governed by an exchange hamiltonian involving, 
besides the actual spin ${\bf S}_r$, also a pseudo-spin-1/2 variable 
${\bf T}_r$, representing the orbital degrees of freedom. 
Unlike the case of a non-degenerate orbital, where antiferromagnetism is 
favored, one finds here a rich phase diagram, depending on the energy of the
two-particle intermediate states available for virtual hopping.\cite{Auerbach}
If Hund's rule coupling is the dominant effect, the lowest doubly 
occupied site is a spin triplet, and spin ferromagnetism turns out to be
favored.\cite{KK} 
Many experimental systems,\cite{KK} including possibly the case of 
TDAE-C$_{60}$,\cite{Auerbach,TDAE:nota} may fit into such a scenario. 
In the present paper we are going to consider a different regime, i.e., 
when the coupling to vibrons is strong enough that an inter-orbital singlet 
state becomes the lowest-energy doubly-occupied virtual state. 
We are going to argue that new interesting phases can show up in this case.
The exchange hamiltonian for this case can be shown to be related
to the following spin-model:\cite{Santoro}
\begin{equation} \label{hst_intro:eqn}
H_{\rm ST} = - J \sum_{<rr'>} 
( 2 {\bf S}_r \cdot {\bf S}_{r'} - \frac{1}{2}) 
( 2 {\bf T}_r \cdot {\bf T}_{r'} - \frac{1}{2}) \;.
\end{equation}
We will show that, in this case, valence bond (VB) states tend to be 
favored, at least in low enough spatial dimensions.

The paper is organized as follows. Section \ref{intro_model:sec} introduces
the hopping model that we will consider. 
In section \ref{spin_ham:sec} we will discuss in detail the hamiltonian
$H_{\rm ST}$ in Eq.\ (\ref{hst_intro:eqn}). 
In section \ref{tst:sec} we study the behavior of the model upon 
introduction of holes, in one dimension. 
Finally, section \ref{conclusion:sec} contains our conclusions, 
and some open issues.

%+++++++++++++++++++++++++++++++++++++++++++++++++++++++++++++++++++++++++
\section{The model} \label{intro_model:sec}
%+++++++++++++++++++++++++++++++++++++++++++++++++++++++++++++++++++++++++

Suppose that at each site $r$ of the lattice we have two degenerate electronic
orbitals and two degenerate phonon modes, of frequency $\omega_0$,
both labeled by an orbital quantum number $l=\pm$. 
Let $c^{\dagger}_{l\sigma}({r})$ be the creation operator for an electron with
spin $\sigma$ and orbital quantum number $l$ at site $r$, and
$b^{\dagger}_l(r)$ the phonon operator. 
In a standard $E\otimes e$ Jahn-Teller problem, the on site
linear electron-phonon interaction term can be written as\cite{Englman}
\begin{equation}
H_{\rm{e-ph}}(r) = \omega_0 \sum_{l=\pm} 
[b^{\dagger}_{l} b_{l} + 1/2] +
(g\omega_0/2) 
\{ T^+ \,[ b^{\dagger}_{-} + b_{+} ] + \rm{H.c.}\} \;,
\end{equation}
where ${\bf T} = (1/2)\sum_{\alpha} \sum_{l,l'}
c^{\dagger}_{l,\alpha} ({\vec \sigma})_{ll'} c_{l',\alpha}$ is 
the orbital (pseudo) spin operator, $\vec\sigma$ are Pauli matrices
and we have omitted the obvious site-label.
We will also consider electron-electron terms, which include a 
Hubbard intra-orbital repulsion $U$, an inter-orbital repulsion $U_{i}$, and
an exchange (Hund's) coupling $J_H$,
\begin{equation}
H_{\rm e-e}(r) = U \sum_{l=\pm} n_{l,\uparrow} n_{l,\downarrow} \,+\,
U_i \sum_{\sigma,\sigma'} n_{+,\sigma} n_{-,\sigma'} \,-\,
J_H {\bf S}_{+} \cdot {\bf S}_{-} \;,
\end{equation}
where ${\bf S}_{l} = (1/2)\sum_{\alpha,\alpha'}
c^{\dagger}_{l,\alpha} ({\vec \sigma})_{\alpha\alpha'} 
c_{l,\alpha'}$ is the spin operator for the $l$--orbital.

Assume now that there are (small) electronic hopping processes between the 
different sites described by a standard tight-binding hamiltonian. 
The total hamiltonian is therefore written as:
\begin{equation} \label{MODEL:eqn}
H = -t \sum_{<rr'>} \sum_{l,\sigma} 
\left[c^{\dagger}_{l\sigma}({r}) c_{l\sigma}({r'}) + \mbox{H.c.}
\right] \,+\, \sum_{r} H_{\rm site}(r) \;,
\end{equation}
where we have neglected overlap integrals between orbitals with different
orbital quantum numbers, and
\begin{equation} 
H_{\rm site}(r) \,=\, H_{\rm e-ph}(r) + H_{\rm e-e}(r) \;.
\end{equation}

We will be interested in the limit in which double occupancy of a site 
is strongly inhibited by the repulsive on-site interactions. 
The low-energy physics is then described by an effective hamiltonian 
in which the double occupancy is only treated as a virtual process to
second-order in the hopping matrix element $t$. 
For a two-fold degenerate orbital, there are six possible two-particle
states to be considered as intermediate states:
an inter-orbital singlet 
$|S\rangle=|S=0,T^z=0\rangle = 2^{-1/2} 
                    (c^{\dagger}_{+,\uparrow} c^{\dagger}_{-,\downarrow} - 
                     c^{\dagger}_{+,\downarrow} c^{\dagger}_{-,\uparrow})
                                                 |0\rangle$,
two intra-orbital singlets
$|{\pm}\rangle=|S=0,T^z=\pm 1\rangle
  =c^{\dagger}_{\pm,\uparrow} c^{\dagger}_{\pm,\downarrow} |0\rangle$,
and an inter-orbital triplet $|T\rangle=|S=1,T=0\rangle$. 
For the case of one electron per site, $n=1$, the resulting effective 
hamiltonian involves two spin-1/2 operators at each site, the spin ${\bf S}$ 
and the orbital pseudo-spin ${\bf T}$. 
The derivation is standard,\cite{Auerbach,Airoldi,Fabrizio} yielding
\begin{eqnarray} \label{general_H:eqn}
H_{\rm eff} &=& \sum_{<rr'>} [ (J_S-J_T+2J_{\pm}) {\bf S}_r \cdot {\bf S}_{r'}
 + (3J_T-J_S) {\bf T}_r \cdot {\bf T}_{r'}
 + 2(J_S-J_{\pm}) T^z_r T^z_{r'} ] \nonumber \\
&& \hspace{5mm} \,+\,
\sum_{<rr'>} [ 4(J_S+J_T) ({\bf S}_r \cdot {\bf S}_{r'})
 ({\bf T}_r \cdot {\bf T}_{r'}) +
 8(J_{\pm}-J_S) ({\bf S}_r \cdot {\bf S}_{r'})
  (T^z_r T^z_{r'}) ] \;.
\end{eqnarray}
The couplings $J_S$, $J_T$, and $J_{\pm}$, generalization of the famous
antiferromagnetic coupling $4t^2/U$ of the non degenerate Hubbard model, 
are essentially given by
$t_{\alpha}^2/\Delta E_{\alpha}$ where $\Delta E_{\alpha}$ 
is the energy difference between a doubly occupied site of type 
$\alpha=S,T,\pm$ and two singly occupied sites, and $t_{\alpha}$ is an
effective hopping matrix element reduced by electron-phonon interactions
(Ham's factor).\cite{Ham}

In absence of electron-phonon coupling ($g=0$), $t_{\alpha}=t$ and
$\Delta E^{(0)}_{S}=U_i+(3/4)J_H$, $\Delta E^{(0)}_{T}=U_i-(1/4)J_H$, 
$\Delta E^{(0)}_{\pm}=U$.\cite{Airoldi,Fabrizio} 
The triplet state is therefore the lowest energy virtual state (under the
realistic assumption $U_i<U$), and $J_T$ the largest coupling in the problem. 
Spin ferromagnetism is the natural outcome of such a scenario.\cite{KK,Auerbach}
The Jahn-Teller coupling can change this picture considerably. 
First of all, standard polaronic effects lead to a decrease of the
energy of the singlet states, primarily the inter-orbital singlet $|S\rangle$, 
with respect to the triplet states $|T\rangle$.\cite{Airoldi,AMT,MTD} 
Second, the effective hopping matrix element $t_{\alpha}$ turns out to be 
larger for the singlet $|S\rangle$ than for the other states.
Ref.\ \cite{Airoldi} gives the results for the $J$'s obtained 
from perturbation theory in $g$, accurate to order $g^4$:
\begin{eqnarray}
J_S &\,=\,& \frac{t'^2 (1+g^4/2)}{U_i+(3/4)J_H-(g^2/2-7g^4/16)\omega_0} \\
J_T &\,=\,& \frac{t'^2}{U_i-(1/4)J_H+(g^2/2-g^4/16)\omega_0}
\nonumber \\
J_{\pm} &\,=\,& \frac{t'^2}{U-(3g^4/16)\omega_0} \nonumber \;.
\end{eqnarray}
In the opposite strong electron-phonon coupling case, $g\to \infty$, 
it was shown in Ref.\ \cite{MTD} that, again, the most important 
doubly-occupied state is the inter-orbital singlet $|S\rangle$, and the model 
can be mapped into a single-band hamiltonian with a spin-1 variable attached 
to each site.\cite{Santoro} 
This also leads to a situation in which, effectively, $J_S$ is the only 
coupling left in the problem.\cite{Santoro}

The mean field phase diagram of the model (\ref{general_H:eqn}) has
been studied in Ref.\ \cite{Auerbach}, together with a few special points
where exact solutions can be obtained in $D=1$. 
As it turns out, there is a region of the phase diagram, not considered so far,
where, as we will argue below, valence bonds (VB) are favored. 
The physical regime in which this region is of relevance is precisely the
JT limit in which $J_S$ is the largest coupling in the problem, i.e., 
the two-particle inter-orbital singlet has lower energy than the corresponding 
triplet. 

More in detail, let us assume that $J_S>J_T$ and define $J=J_S-J_T$. 
In a bipartite lattice the effective hamiltonian, after performing the 
canonical transformation ${\bf T}^{\pm} \to (-1)^r {\bf T}^{\pm}_r$, can be 
written as: 
\begin{equation}
H_{\rm eff} = H_{\rm ST} + \Delta H
\end{equation}
where
\begin{equation} \label{ST_1:eqn}
H_{\rm ST} \,=\, J \sum_{<rr'>} \left[ {\bf S}_r \cdot {\bf S}_{r'}
 + {\bf T}_r \cdot {\bf T}_{r'}
 - 4 ({\bf S}_r \cdot {\bf S}_{r'}) ({\bf T}_r \cdot {\bf T}_{r'}) \right] \;,
\end{equation}
and
\begin{eqnarray} \label{Delta_H:eqn}
\Delta H &\,=\,& 4J_T \sum_{<rr'>} \left[ T^z_r T^z_{r'} 
 - (2 {\bf S}_r \cdot {\bf S}_{r'} +1/2) \;
{\bf T}^{\perp}_r \cdot {\bf T}^{\perp}_{r'} \right] \nonumber \\
&& \,+ 4J_{\pm} \sum_{<rr'>} (2 {\bf S}_r \cdot {\bf S}_{r'} -1/2) 
(T^z_r T^z_{r'} + 1/4) \;.
\end{eqnarray}

In the next section we will concentrate on $H_{ST}$, showing first of all
that it certainly has a VB-solid ground state in $D=1$, and suggesting that 
the same might be true in $D=2$. 
The presence of a gap in the excitation spectrum implies the robustness of 
such a phase with respect to a small enough perturbation $\Delta H$. 

Next, we will consider the question of the behavior of a valence bond 
state upon doping. The resulting generalization of the $t-J$ model will 
be discussed, for the $D=1$ case, in section \ref{tst:sec}.

%+++++++++++++++++++++++++++++++++++++++++++++++++++++++++++++++++++++++++
\section{One electron per site: The spin hamiltonian.} \label{spin_ham:sec}
%+++++++++++++++++++++++++++++++++++++++++++++++++++++++++++++++++++++++++

As previously discussed, when there is exactly one electron per site 
($n=1$) and one assumes the inter-orbital singlet $|S\rangle$ to be
the only important intermediate two-particle state, the exchange hamiltonian
governing the spin and orbital degrees of freedom can be written as
\begin{equation} \label{ST_2:eqn}
H_{\rm ST} = - J \sum_{<rr'>} 
( 2 {\bf S}_r \cdot {\bf S}_{r'} - \frac{1}{2}) 
( 2 {\bf T}_r \cdot {\bf T}_{r'} - \frac{1}{2}) \,=\,
- J \sum_{<rr'>} (P^{(S)}_{rr'}-1)(P^{(T)}_{rr'}-1)  \;,
\end{equation}
where $P^{(S)}_{rr'}=2 {\bf S}_r \cdot {\bf S}_{r'} + 1/2$ 
is a permutator between the spins at sites $r$ and $r'$, 
and similarly $P^{(T)}$ for the pseudo-spins.
This form will be particularly useful in the following. 

This model shows an obvious spin and pseudospin rotation invariance
described by the $SU(2)\otimes SU(2)$ group. However, the full 
symmetry group displayed by (\ref{ST_2:eqn}) is much larger and 
includes $SU(4)$. Indeed, one can show that, in a bipartite lattice,
$H_{\rm ST}$ commutes with the 15 operators
$S^{\alpha}=\sum_r S^{\alpha}_r$, $T^{\alpha}=\sum_r T^{\alpha}_r$, 
and $L^{\mu\nu}=2\sum_r (-1)^r S^{\mu}_r T^{\nu}_r$ which generate the
Lie algebra $su(4)$ \cite{SU4}.
Due to the staggering in the $L^{\mu\nu}$ operators, these generators 
{\em do not} commute with the spatial symmetries which interchange the 
two sublattices. 
Therefore, the full symmetry group is even larger than the 
aforementioned $SU(4)$. 
Finally, we notice that this hamiltonian differs from the exactly soluble 
$SU(4)$ Sutherland model $H=J \sum_{<rr'>} P^{(S)}_{rr'}P^{(T)}_{rr'}$ 
obtained by a natural generalization of the Heisenberg hamiltonian. 
Rather, it is possible to show that the $SU(n)$-invariant model introduced by 
Affleck \cite{Affleck} 
\begin{equation}
H_n = \frac{1}{n} \sum_{<rr'>} S^{\alpha}_{\beta}(r) 
\bar{S}^{\beta}_{\alpha}(r') \;,
\end{equation}
is unitarily equivalent to $H_{\rm ST}/4J$ for $n=4$. 
Here $S^{\alpha}_{\beta}$ and $\bar{S}^{\beta}_{\alpha}$ generate, respectively,
the fundamental and anti-fundamental representations of $SU(n)$.\cite{Affleck}

In $D=1$, Affleck has shown that $H_n$, in the limit $n\to \infty$, has
two simple VB-solid ground states, illustrated in Fig.\ \ref{vbgs:fig}. 
%----------------------- MODIFIED SENTENCE -----------------------------------
Moreover, the $n=3$ case has also been shown to have a dimerized ground 
state, although with a large spin-spin correlation length.\cite{Affleck_n=3} 
As the tendency to dimerization is likely to increase with $n$,\cite{Affleck} 
we expect that the $n=4$ case, of interest to us, will also have a 
VB-solid type of ground state. 
This expectation will be explicitly shown to be correct in $D=1$. 
%-----------------------------------------------------------------------------

Let us first show that singlet bonds between $S$ and $T$ variables are
very natural objects to introduce in our problem, for any $D$. 
If we denote by 
$(rr')_S=2^{-1/2}(\uparrow_r\downarrow_{r'}-\downarrow_r\uparrow_{r'})_S$
a singlet bond formed between the spins at $r$ and $r'$, 
it is easy to show that \cite{Anderson}
\begin{eqnarray}
&& (P^{(S)}_{rr'}-1) (rr')_S = -2 (rr')_S \nonumber \\
&& (P^{(S)}_{r'l}-1) (rr')_S (ll')_S = (r'l)_S (rl')_S \;.
\end{eqnarray}
Similar results apply to $P^{(T)}-1$ when acting on valence bonds formed from
the pseudo-spin variables, $(rr')_T$. 
If we denote by $(rr')=(rr')_S (rr')_T$ the product of singlet bonds between
$r$ and $r'$ for both $S$ and $T$, and by 
$Q_{rr'} = [P^{(S)}_{rr'}-1] [P^{(T)}_{rr'}-1]$, we find that 
$H_{\rm ST}=-J\sum_{<rr'>}Q_{rr'}$ and
\begin{eqnarray} \label{Qaction:eqn}
&& Q_{rr'} (rr') = 4 (rr') \nonumber \\
&& Q_{r'l} (rr') (ll') = (r'l) (rl') \;.
\end{eqnarray}
First observation: the subspace of valence bond states 
in which bonds are simultaneously S- and T-singlets is left invariant 
by the hamiltonian. 
(In all figures, we will denote by a line such simultaneous S- and T-singlets.)
Such a subspace is, however, still overcomplete.\cite{Pauling}
In $D=1$, nevertheless, if we apply $H_{\rm ST}$ any number of times to the
valence bond solid in Fig.\ \ref{vbgs:fig}, we generate a subset of valence 
bond configurations which are linearly independent (although not orthogonal).
These are the valence bond configurations associated to non-crossing Lewis
diagrams.\cite{Pauling} 
It is known that such a basis can be conveniently used to work out efficient 
exact diagonalization algorithms in small chains.\cite{Soos} 
The ground state wave function for a chain of $N=8$ sites, for instance, 
can be easily obtained by solving a simple $3\times 3$ problem. 
Fig.\ \ref{8siti:fig} shows the result obtained for the ground state wave 
function, and the lowest excited state of momentum $\pi$, in terms of 
VB configurations. 
 
Second observation: as is apparent from Eq.\ \ref{Qaction:eqn}, 
a nearest-neighbor bond contributes a diagonal energy of $-(4J)$, 
whereas the off-diagonal matrix elements connecting different VB configurations
are smaller by a factor $4$, i.e., $-J$. 
For the $SU(n)$-model of Affleck, the off-diagonal matrix element are smaller by
a factor $1/n$, which implies that only diagonal energies are 
retained in the limit $n\to \infty$. 
The ground states, for $n=\infty$, are therefore the VB configurations 
with the maximum number ($=N_{\rm sites}/2$) of nearest-neighbor bonds 
connected by VBs. 
In $D=1$, there are only two such states, i.e., the VB-solids in 
Fig.\ \ref{vbgs:fig}.

To check if the VB scenario remains correct for our $n=4$ case,
we resort to exact Lanczos diagonalizations of chains up to $14$ sites, and 
to Green Function Monte Carlo for longer chains. 
Fig.\ \ref{gap1d:fig} shows the finite-size value of the gap between the 
ground state and the first excited state both in the $S^z_{TOT}=T^z_{TOT}=1$ 
($\Box$) and in the $S^z_{TOT}=T^z_{TOT}=0$ sectors ($\bigcirc$). 
The full symbols refer to $H_{\rm ST}$, whereas the open symbols refer to
two decoupled Heisenberg chains (no quartic term in Eq.\ \ref{ST_1:eqn}). 
All these excited states have momentum $\pi$ relative to the ground state.
For the Heisenberg case, the excited state is a triplet and the model has 
gapless excitations with gaps decreasing as $1/L$ for 
$L\rightarrow \infty$.\cite{sizescal}
For $H_{\rm ST}$, the situation is different: The lowest excited state is
a singlet, with a finite size gap $\Delta E(S=0)$ decreasing faster than $1/L$ 
for $L\rightarrow \infty$. 
The triplet, instead, lies above with a gap $\Delta E(S=1)$ extrapolating 
to a finite value as $L\rightarrow \infty$. 
This is a clear signal of a {\em degenerate\/} infinite volume ground 
state with a {\em spontaneous breaking of translational symmetry\/} and a gap
to all excitations, consistent with a VB phase. 
Further support to this interpretation can be gained by direct inspection of
the spin-spin correlations, $\langle S^z_i S^z_j \rangle$, the
dimer-dimer correlations along a chain 
$\langle (S^z_iS^z_{i+1}) (S^z_jS^z_{j+1}) \rangle$, 
and on different chains 
$\langle (S^z_iS^z_{i+1}) (T^z_jT^z_{j+1}) \rangle$. 
Fig.\ \ref{sq1d:fig} shows a log-log plot of the size scaling of the peak of 
the different structure factors, located at momentum $\pi$. 
While the dimer-dimer structure factor peaks, 
$\sum_i (-1)^i \langle (S^z_0S^z_{1}) (S^z_iS^z_{i+1}) \rangle$ and
$\sum_i (-1)^i \langle (S^z_0S^z_{1}) (T^z_iT^z_{i+1}) \rangle$, 
tend to diverge linearly with the length $L$ of the chain, the usual spin-spin 
structure factor tends to a finite value in the thermodynamic limit.

Finally, simple variational arguments also point to a VB ground state in 
$D=1$, while being, we believe, not conclusive in $D=2$. 
A VB phase benefits highly from the large values of 
$\langle {\bf S}_r \cdot {\bf S}_{r'} \rangle = 
\langle {\bf T}_r \cdot {\bf T}_{r'} \rangle = -3/4$ 
on the singlet bonds, due to the presence of the ``square'' of the 
singlet contribution in the quartic term of $H_{\rm ST}$, which enters with
a factor $4$.
Let us compare, for instance, the energy of the following two states: 
(i) The product of two independent Heisenberg ground states,
$|\Psi_{\rm 2-H}\rangle = |\Psi^{(\sigma)}_{\rm H}\rangle \otimes
                          |\Psi^{(\tau)}_{\rm H}\rangle$, 
and (ii) the product of two VB states,  
$|\Psi_{\rm 2-VB}\rangle = |\Psi^{(\sigma)}_{\rm VB}\rangle \otimes
                           |\Psi^{(\tau)}_{\rm VB}\rangle$ 
where each VB state has the simple form of a product of dimers on 
adjacent sites. In $D=1$ we take 
\begin{equation}
|\Psi_{\rm VB}\rangle = (1 2) \, (3 4) \, \cdots \, (L-1 L) \;,
\end{equation}
where $(ij) = (\uparrow\downarrow-\downarrow\uparrow)/\sqrt{2}$ 
denotes a singlet between sites $i$ and $j$. 
In $D=2$ we consider the simplest short range VB state: a 
columnar dimer state, or, equivalently, any state with nearest-neighbor 
pairs coupled into singlets in an arbitrary manner (there is a huge 
degeneracy). 
The energy per site of the two states are:
\begin{eqnarray}  \label{E_2H_2VB:eqn}
\epsilon_{\rm 2-H} &=& 2 \epsilon_{\rm H} - 
4J D(\epsilon_{\rm H}/DJ)^2 \;, \nonumber \\
\epsilon_{\rm 2-VB} &=& 2 \epsilon_{\rm VB} - 4J (1/2)(-3/4)^2 \;.
\end{eqnarray}  
Here $\epsilon_{\rm H}/J=D\langle {\bf S}_r \cdot {\bf S}_{r'} \rangle$ 
($r$ and $r'$ nearest neighbors) and $\epsilon_{\rm VB}=(-3/8)J$
are the energy per site of the Heisenberg ground state and of the VB
state, respectively.
For the Heisenberg model we have 
$\epsilon_{\rm H}/J=(-\ln{2}+1/4)\approx-0.4431$ (in $D=1$) and 
$\epsilon_{\rm H}/J=\approx-0.66$ (in $D=2$), whereas for the crude VB state 
considered here, the expression for $\epsilon_{\rm 2-VB}$ 
turns out to be independent of $D$.
The factor $(1/2)$ in the expression for $\epsilon_{\rm 2-VB}$ is due to the
fact that only {\rm half\/} of the bonds in the VB state enjoy the large 
singlet-singlet pairing $-4J(3/4)^2$. 
In spite of this reduction, the coefficient of the quartic term is substantially
larger for the VB state than for the Heisenberg state in both $D=1$ and $D=2$ 
($\approx-0.2812 (4J)$ for the VB state, to be compared to 
$\approx-0.1964 (4J)$ and $\approx-0.218 (4J)$ for the $D=1$ and $D=2$ 
Heisenberg, respectively). 
In $D=1$ the VB state wins over the Heisenberg state. 
Generally speaking, an increase in the coordination number tends to stabilize 
a N\'eel-like antiferromagnet with respect to VB states. 
In $D=2$, the previous crude variational estimate would indeed give the 
Heisenberg state as favored. 
The VB state considered here is, however, very poor: for instance, 
its energy per site, neglecting the quartic term, is only $-0.375 J$, 
whereas it is well known that short-range RVB states close in energy to 
the Heisenberg ground state can be constructed.\cite{Doucot}
As a consequence, we believe the present variational estimate to be
not conclusive in $D=2$. 

In case the VB picture should be energetically favored, we face the problem of 
determining the possible orderings of the dimers in the lattice. In 
principle, either a VB crystal with broken translational symmetry
or a homogeneous spin liquid may occur.
Following Affleck's $D=1$ approach, \cite{Affleck} a possible way of
tackling this problem, in $D=2$, is from the $n\to \infty$ limit. 
For $n=\infty$, once again, all the states made up of the
maximum possible number ($=N_{\rm sites}/2$) of nearest-neighbor S- and T-VBs 
-- we refer to them as NNVB configurations -- are degenerate, linearly
independent (although not orthogonal) ground states. 
Contrary to the $D=1$ case, where only two such states exist 
(see Fig.\ \ref{vbgs:fig}), in $D=2$ this implies a huge degeneracy.
For $n\to \infty$, however, there are off-diagonal matrix elements connecting 
different NNVB states (smaller by a factor $1/n$, as previously discussed),  
which lift this degeneracy to first order. 
It is straightforward to verify, using Eq.\ \ref{Qaction:eqn}, that the 
effective hamiltonian within the subspace of NNVB states 
(to first order in $1/n$)
is just the quantum dimer model (QDM) first introduced in Ref.\ \cite{Rok_Kiv}
\begin{equation}
H_{QMD} = -J \sum_{\rm plaquette} 
\left\{ | \, = \; \rangle \langle \; || \; | + {\rm H.c.} \right\} 
\,+\, V \sum_{\rm plaquette} 
\left\{ | \, =  \; \rangle \langle \, =  \; | + 
        | \; || \; \rangle \langle \; || \; |  \right\} \;,
\end{equation}
for the particular case of $V=0$. (Here $||$ and $=$ schematically indicate
adjacent S- and T-VBs in the y and x direction.)
The QDM is known to have a {\em columnar solid\/} phase in a region of 
parameter space close to $V=0$.\cite{Rok_Kiv,Sachdev} 
Recently, an exact diagonalization study of the QDM has revealed that 
such a columnar solid phase has, more precisely, the features of a 
{\rm plaquette RVB\/}, i.e.\ translational invariance in one direction is 
spontaneously broken, but the spin-Peierls ``columns'' are disordered as a 
consequence of a resonance of parallel bonds in each plaquette.\cite{Leung}
This picture would yield a fourfold degenerate ground state in two dimensions. 

We conclude that, for $n\to \infty$, our model has a columnar VB phase
in $D=2$. Given the fact that our crude variational calculation appears
to favor a N\'eel state in $D=2$, we believe that the issue of the actual 
behavior of our $n=4$-case, $H_{\rm ST}$, in $D=2$ is still open. 
This issue will be addressed in a future numerical study.\cite{Guidoni}

%+++++++++++++++++++++++++++++++++++++++++++++++++++++++++++++++++++++++++
\section{Less than half-filling.} \label{tst:sec}
%+++++++++++++++++++++++++++++++++++++++++++++++++++++++++++++++++++++++++

The motion of holes in a VB state has been the subject of considerable
research starting from the early works on high temperature superconductivity
\cite{KRS}. Now, we are able to investigate this problem on the basis of a 
consistent microscopic model which does have a VB solid as a ground state.
This study may shed light on the issue whether holons and spinons
decouple in VB states and whether the VB spin background favors effective
hole-hole attraction. Therefore, we analyze in some detail 
the case of filling $n<1$ in one dimension.
For large repulsive interactions (small $J$) the effective hamiltonian 
describing the system is simply the analog of the $t-J$ model
\begin{equation} \label{htst:eqn}
H_{t-ST} = P_o K P_o - 4J\sum_{<rr'>} 
n_r n_{r'} ({\bf S}_r \cdot {\bf S}_{r'} - \frac{1}{4}) 
(-{\bf T}_r \cdot {\bf T}_{r'} + 2 {T}^z_r {T}^z_{r'} 
- \frac{1}{4}) + ({\rm 3-site \,terms}) \;,
\end{equation}
where $K$ is the hopping part of the original hamiltonian 
(Eq.\ \ref{MODEL:eqn}), $P_o$ is the projector onto states without double 
occupancy, and $n_r$ is the occupation number operator. 
In $D=1$, analogously to the Hubbard case, \cite{ogata_shiba,factorize} 
the (real-space) wavefunction {\em factorizes\/}, for $J\to 0$, 
into a spinless fermion part times the wavefunction of a suitable spin 
hamiltonian $H_{\rm spin}$ on the {\em squeezed chain\/} \cite{squeez}
\begin{equation} \label{factor}
\Psi(x_1,\cdots,x_N;\sigma_1,\cdots,\sigma_N,\tau_1\cdots,\tau_N) = 
\psi_{SF}(x_1,\cdots,x_N) 
\Phi(\sigma_1,\cdots,\sigma_N,\tau_1\cdots,\tau_N) \;. 
\end{equation}
The position of the electrons is determined by the spinless fermion 
wavefunction $\psi_{SF}(x_1,\cdots,x_N)$, while the spin and orbital
ordering is governed by a spin hamiltonian which is obtained by taking 
the average of $H_{t-ST}$ over the spinless fermion state.
The result is simply related, by an overall change of energy scale and a
canonical transformation, to the previous spin model $H_{\rm ST}$ 
(Eq.\ \ref{ST_2:eqn}): 
\begin{equation} \label{hspin:eqn}
H_{\rm spin} = -4J n[1-\frac{\sin{(2\pi n)} }{2\pi n}] \sum_{<ij>} 
({\bf S}_i \cdot {\bf S}_{j} - \frac{1}{4}) 
(-{\bf T}_i \cdot {\bf T}_{j} + 2 T^z_i T^z_j - \frac{1}{4}) \;.
\end{equation}
The physical origin of this factorization relies on a
general feature of one dimensional models with nearest neighbor 
hopping: In the $U\to\infty$ limit, the spatial ordering of the 
particles is conserved by the dynamical evolution, irrespectively 
of the spin configuration. Therefore, the hamiltonian can be diagonalized
in each subspace defined by a given spin ordering and the result does not
depend on the chosen spin configuration. The effective hamiltonian
governing the charge degrees of freedom is that of a free spinless fermion 
gas (with suitable boundary conditions which do depend on the spin
wavefunction \cite{sandro_alberto}) whose eigenfunctions are $\psi_{SF}$.
The degeneracy among different spin configurations is lifted to $O(J)$
by the magnetic term in $H_{t-ST}$ (\ref{htst:eqn}), leading to the complete
classification (\ref{factor}) of the eigenstates of $H_{t-ST}$ to 
lowest order in $J$.

The spin hamiltonian in Eq.\ \ref{hspin:eqn}, we argued before, has a spin gap. 
Therefore, we conclude that, while the charge sector is gapless (metallic) for 
large positive $\Delta E_S$ and $n\ne 1$, {\em the spin sector is gapped at all 
densities\/} and has a doubly degenerate ground state in the thermodynamic
limit. 
However, such a spin gap is not related to the formation of bound electron 
pairs. In fact, the charge degrees of freedom behave as free particles with 
no tendency towards pairing: the superconductive correlations are 
not enhanced in the ground state and decay as $\sim x^{-2}$.
This state can be rather interpreted as the one dimensional realization of 
an ``itinerant''  valence bond state where spinons bind, leaving holons in a 
Luttinger Liquid state. 
(The situation is quite similar to the strong coupling phase of the 
$t-t'$ Hubbard model in $D=1$, for $t'>t/2$.\cite{Michele}) 
The very presence of such a phase is then a remarkable consequence of 
one dimensional spin--charge decoupling.

The factorization of the wavefunction at $J\to 0$ can be used to calculate 
various physical quantities of this spin-gapped metallic phase, following 
analogous calculations in the Hubbard case.\cite{ogata_shiba,sandro_alberto}
In particular, density correlation functions just coincide with 
those of a spinless fermion gas, showing power law decay at large distances
($\sim x^{-2}$)
while the spin-spin (and pseudospin) correlation still decays exponentially,
similarly to the half filled case. 
Another interesting property which can be evaluated is the one-hole 
Green's function providing a quantitative description of the motion of one 
hole in a short range VB state.
The picture of a static VB crystal would yield an exponentially decaying 
Green's function in real space, but 1D fluctuations severely affect the
asymptotic form of the Green's function. 
The calculation can be carried out exactly in the $J\to 0$ limit, where the 
factorization property of all eigenfunctions holds. 
In fact, in this limit, the energy levels of a single hole in a VB state 
coincide with those of a free particle in vacuum plus $O(J)$ corrections 
depending on the energy of the spin background. 
Moreover, the energy, to lowest order, just depends on the 
{\em holon momentum}, i.e. on the momentum $k$ of the spinless fermion part 
of the wavefunction $\psi_{SF}$: $E(k)=2t\cos k + O(J)$. 
On the other hand, the holon momentum is related to the total momentum 
$p$ of the state by the conservation law: $p=k+Q$. 
Here $Q$ is the momentum of the spin part of the
wavefunction $\Phi$ and is quantized in units of $2\pi/(L-1)$, the
spin model being defined on the squeezed chain of $L-1$ sites.
Therefore, for $J\to 0$, the one-hole energy levels of the model, at a given 
total momentum $p$, can be written in terms of the spinon momentum: 
$E_{p,Q}=2\cos (p-Q)$, i.e., of the momentum of the spin part of the 
factorized wavefunction. 
The corresponding density of states coincides with that of a single free 
fermion:
\begin{equation}
N(\omega)={1\over 2\pi} {1\over \sqrt{4t^2-\omega^2}} \;,
\end{equation}
with singularities at the band edges $\omega=\pm 2t$. 
These singularities are due to the quadratic dependence of the energy spectrum
on the holon momentum which will be probably cut-off by the 
presence of spin excitations to $O(J)$.

The calculation of the single hole Green's function
\begin{equation} \label{holegreen}
G(p,t)=
i\, \langle \Psi | \,c^{\dagger}_{p,\sigma} \,e^{-i t (H-E_0-i\delta)}
                   \,c_{p,\sigma}\, |\Psi \rangle \,\theta(t)
\end{equation}
is conveniently carried out by use of the Lehmann representation of the
corresponding spectral function:
\begin{eqnarray} \label{spectral}
A(p,\omega) \,=\, {1\over\pi} {\rm Im}\; G(p,\omega) \,&=&\, 
\sum_s | \langle \Psi_{s}^{h}\,| \, c_{p,\sigma}\, |\,
\Psi \rangle |^2  \; \delta (\omega + E_0 - E_s) \nonumber\\ 
&=& \sum_{Q} Z(Q) \, \delta (\omega - E_{p,Q}) \;,
\end{eqnarray}
where, in the last equation, use has been made of the special form of the
energy spectrum as $J\to 0$. 
The usefulness of this representation rests on the knowledge of the weight 
function $Z(Q)$ which, following Ref. \cite{sandro_alberto}, can be written 
in terms of a non-local correlation function involving only spin 
(and pseudospin) variables:
\begin{eqnarray} \label{Z:eqn}
Z(Q)\,&=&\, {1\over 2} {1 \over L-1} \sum_{j=0}^{L-2} e^{-iQj} \Omega(j) 
\nonumber \\
\Omega(j) \,&=&\, \langle \Phi\, | \, 
(2 {\bf T}_j\cdot {\bf T}_{j-1} +{1\over 2})
(2 {\bf S}_{j}\cdot {\bf S}_{j-1} +{1\over 2})
\dots
(2 {\bf T}_1\cdot {\bf T}_0 +{1\over 2})
(2 {\bf S}_1\cdot {\bf S}_0 +{1\over 2})
\, |\, \Phi \rangle \;.
\end{eqnarray}
Such a correlation function can be evaluated on the ground state of the
$H_{\rm ST}$ hamiltonian by Lanczos diagonalization and, similarly 
to the Heisenberg model case, shows negligible size dependence. 
Calculations have been carried out up to $L=18$ sites, and the results 
are shown in Fig.\ \ref{Z:fig}. 
A clear singularity in the quasiparticle weight $Z(Q)$ is visible at the 
spinon Fermi momentum $Q_F= \pi/2$. 
The analysis of Lanczos data suggests a power law divergence as
$Q\to Q_F^{\pm}$, in close similarity to the analogous behavior encountered 
in the {\em gapless} Heisenberg model. A quantitative determination of
the critical exponent is however precluded by the severe limitation in system
size associated to the Lanczos method. 
In the thermodynamic limit, this singularity gives rise to branch cuts 
in the analytic structure of the spectral function (\ref{spectral}):
\begin{equation} \label{thermo}
A(p,\omega)\,=\, (L-1) \left[ Z(Q_{+}(p,\omega))+  Z(Q_{-}(p,\omega)) \right]
N(\omega) 
\end{equation}
along the lines defined by $\vert Q_{\pm} \vert=\pi/2$, where 
$Q_{\pm}(p,\omega)\,=\, p \pm \arccos ({\omega \over 2})$ i.e. 
at $\omega_p=\pm2\sin p$. The form of this dispersion relation suggests 
nearest neighbor hopping processes of the hole across the lattice.

The overall picture of hole motion in a Valence Bond state is therefore 
quite similar to the case of holes in one dimensional quantum antiferromagnets, 
being characterized by power law tails in the single particle Green's function
at the dispersion energy $\omega_p$. 
This behavior can be traced back to the presence of branches of {\em gapless} 
collective modes above the spin gap which allow the decay of the
spin $1/2$ excitation created by the hole motion. 
These features of hole propagation should be 
contrasted to the properties of charges in Ising 
antiferromagnets, where the excitation spectrum is discrete and the 
band is dispersionless.

%+++++++++++++++++++++++++++++++++++++++++++++++++++++++++++++++++++++++++
\section{Conclusions} \label{conclusion:sec}
%+++++++++++++++++++++++++++++++++++++++++++++++++++++++++++++++++++++++++

In this paper we have discussed in some detail a hopping model of 
strongly repelling electrons in the presence of local orbital degeneracies 
not removed by crystal field effects. 

For the insulating case (one electron per site and large Coulomb repulsions) 
we have studied a particular exchange hamiltonian, Eq.\ (\ref{hst_intro:eqn}). 
It describes the low-energy physics of this system in the regime in which the 
inter-orbital singlet is the lowest-energy intermediate state available to
hopping. 
This regime tends to be further favored by a relatively strong electron-phonon
coupling. 
At variance with the standard Hund's rule case (the high spin two-particle state
has lowest energy), which would instead favor spin ferromagnetism, we find in
our case a strong tendency to the formation of a statically ordered
Valence Bond (VB) phase. 
In $D=1$, this has been firmly established by using variational
arguments, a large-$n$ limit, and exact diagonalization results.
Given the presence of a sizeable spin-gap, weakly coupled chains should 
also have a VB ground state. 

More uncertain is, instead, the outcome in the isotropic $D=2$ case,
where a large-$n$ approach would predict a columnar VB phase, whereas
a crude variational calculation rather points to a N\'eel-like antiferromagnet.
A numerical study, presently under way, will clarify, we hope, this matter.

Finally, we have considered the behavior of the system upon hole
doping in the $D=1$ case. The factorization of the wavefunctions, valid in
the strong repulsive limit and peculiar to the $D=1$ case, is the crucial 
ingredient used to tackle the problem. The outcome is a kind of ``itinerant''
VB state, where charge degrees of freedom are gapless, but the spin and
orbital degrees of freedom are gapped and VB-like. 

The motivation for our study has been primarily theoretical. 
Originally inspired by dynamical Jahn-Teller molecular lattices, such as
$C_{60}$ compounds, the ingredients used are, admittedly, extremely idealized, 
and not easily applicable to an existing realistic situation. 
In a real JT lattice, both the crystal field effect, and the 
intermolecular elastic coupling cannot normally be neglected, and may favor
a cooperative static JT distortion of all molecules.\cite{Englman}
Second, intermolecular electron hopping tends to be comparable, or even 
larger in comparison with relevant vibron energies. 
Apart from this, realistic hopping matrix elements have a strong orientational 
dependence in more than one dimension, and, in general, there is no reason 
for neglecting overlap integrals between orbitals with 
different orbital quantum numbers (see Eq.\ (\ref{MODEL:eqn})). 
Third, a detailed description of the structure of the doubly occupied states
at each site brings in correction terms to the exchange 
hamiltonian which turns out to be, in general, strongly asymmetric in orbital 
space (see Eq.\ (\ref{Delta_H:eqn})). 
 
On the opposite side, the exchange hamiltonian we have studied is elegant,
parameter-free and prototypical of strong correlations in an orbitally
degenerate lattice. 
The physics uncovered has a definite robustness against ``small'' 
perturbations, due to the gap present in the excitation spectrum. 
The VB gapped state has a clear resemblance to a spin-Peierls state
in $D=1$. The spin-gapped metal found upon doping is interesting.
The basic ingredients needed for a realistic system to be a candidate VB 
described in this work are:
i) orbital degeneracy not trivially removed by cooperative JT and/or
crystal field effects;
ii) relatively large molecules with a strong electron-phonon coupling, 
so as to make the inter-orbital singlet favored as compared to the Hund's 
rule triplet;
iii) narrow bands with relatively large on-site Hubbard $U$, so as to stabilize
a Mott-Hubbard insulator at exactly one electron per site;
iv) reduced dimensionality. 

$C_{60}$ charge transfer compounds, based on $C_{60}^-$ ions,\cite{C60:rev} 
are potential candidates to the realization of such a scenario.
$C_{60}$ has a triply degenerate $t_{1u}$ molecular orbital, coupled 
to several intramolecular vibrations, resulting in an important 
dynamical JT effect.\cite{AMT} 
The electron-phonon coupling leads to a substantial pairing energy 
$\approx 0.1 eV$ which is, however, overwon by a substantially larger 
Hubbard $U\approx 1-1.5 eV$,\cite{Gunnarsson} resulting in a Mott insulating 
behavior. 
The lattices tend to have anisotropic lattice constants, with a
pronounced quasi-one-dimensional character.\cite{C60:rev} 
As for the virtual intermediate states of $C_{60}^{2-}$, singlets and triplet
are close in energy, at least in solutions.\cite{C60:exp} 
If the triplet prevails, as is possibly the case in
TDAE-$C_{60}$,\cite{Auerbach,TDAE:nota} there can be spin ferromagnetism. 
We believe, however, there is room in some future system for the alternative 
possibility of a spin gap, and the formation of the spin-orbital VB state 
described here.

ACKNOWLEDGMENTS --
It is a pleasure to thank S. Sorella and M. Fabrizio for many stimulating
discussions. AP gratefully acknowledges hospitality at SISSA. 
Work at SISSA and ICTP was co-sponsored by INFM project HTSC, and
by EU contract ERBCHRXCT940348. 

%%%%%%%%%%%%%%%%%%%%%%%%%%%%%%%%%%%%%%%%%%%%%%%%%%%%%%%%%%%%%%%%%%%%%%%%%
%                               FIGURES
%%%%%%%%%%%%%%%%%%%%%%%%%%%%%%%%%%%%%%%%%%%%%%%%%%%%%%%%%%%%%%%%%%%%%%%%%
\newpage
\begin{center}
{\bf FIGURE CAPTIONS}
\end{center}

\begin{figure} 
\caption{The two VB solid configurations in $D=1$.}
\label{vbgs:fig}
\end{figure} 

\begin{figure} 
\caption{The ground state (momentum $0$, $+$) and first excited state 
(momentum $\pi$, $-$) for $H_{\rm ST}$ on a chain of $8$ sites in terms
of S- and T-VB configurations. ($\alpha=0.566\cdots$, $\beta=0.416\cdots$) and
($\alpha=0.268\cdots$, $\beta=0.102\cdots$), respectively, 
for the two states. }
\label{8siti:fig}
\end{figure} 
  
\begin{figure} 
\caption{Finite size gaps for the lowest excited states of the hamiltonian 
$H_{\rm ST}$ in $D=1$  (solid symbols) compared to the Heisenberg chain case 
(open squares). The data are obtained by exact diagonalization, for
chains up to $14$ sites, and by Green Function Monte Carlo for longer chains. 
The excited state for the Heisenberg case is a triplet and 
$\Delta E/J=\pi^2/(2L)+\cdots$, shown by the dashed line. 
The solid straight line, extrapolating to a finite value for $L\to \infty$
($\approx 0.5969$), is obtained by a quadratic least-square fit to the triplet
excitation results. This figure shows that, whereas the triplet excitations are
gapped, the singlet gap decreases faster than $1/L$, signalling spontaneous
symmetry breaking, consistent with a VB ground state.}
\label{gap1d:fig}
\end{figure} 

\begin{figure} 
\caption{Log-log plot of the $\pi$-component of the Fourier transform of 
different correlation functions for the hamiltonian $H_{\rm ST}$ in $D=1$. 
Open squares refer to $\langle S^z_i S^z_j\rangle$, open and solid circles
refer to 
$\langle (S^z_i S^z_{i+1}) (S^z_j S^z_{j+1}) \rangle$ and
$\langle (S^z_i S^z_{i+1}) (T^z_j T^z_{j+1}) \rangle$, respectively. 
The dashed line has slope $1$, for comparison. The dimer-dimer correlations
grow fastest with size, again consistent with a VB ground state.}
\label{sq1d:fig}
\end{figure} 

\begin{figure} 
\caption{The function $Z(Q)$ defined, in Eq.\ \protect{\ref{Z:eqn}}, 
determining the single-hole spectral function in the thermodynamic limit 
(see Eq.\ \protect{\ref{thermo}}). 
The data are obtained from exact diagonalizations of chains up to $L=18$.  }
\label{Z:fig}
\end{figure}

%%%%%%%%%%%%%%%%%%%%%%%%%%%%%%%%%%%%%%%%%%%%%%%%%%%%%%%%%%%%%%%%%%%%%%%%%
%                               BIBLIOGRAPHY
%%%%%%%%%%%%%%%%%%%%%%%%%%%%%%%%%%%%%%%%%%%%%%%%%%%%%%%%%%%%%%%%%%%%%%%%%


\begin{thebibliography}{99}
\bibitem{factorize}
Since we do not have a Bethe-Ansatz solution, we have tested this conclusion
numerically from exact diagonalization data.
\bibitem{squeez} 
By {\em squeezed chain} we mean the ring
defined by the sites occupied by the electrons. Clearly the squeezed
chain is defined for every {\em fixed} real space configuration of the 
electrons.
\end{thebibliography}
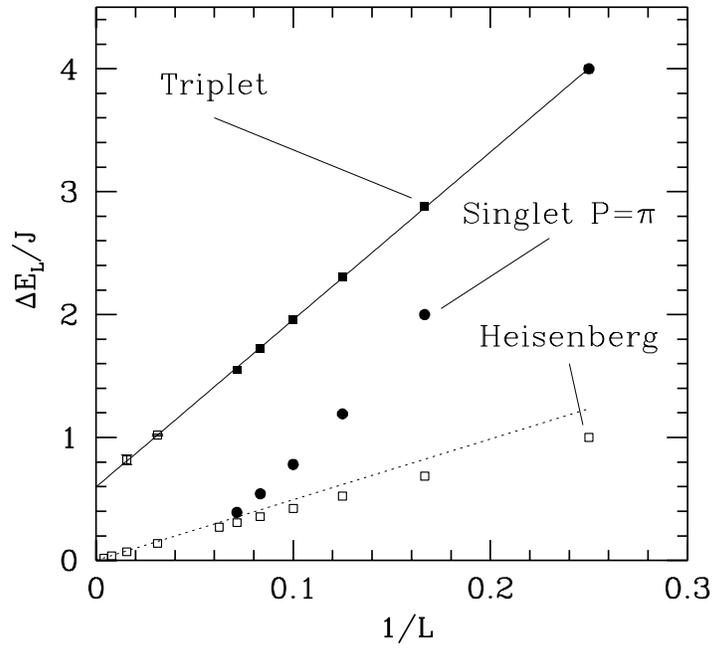
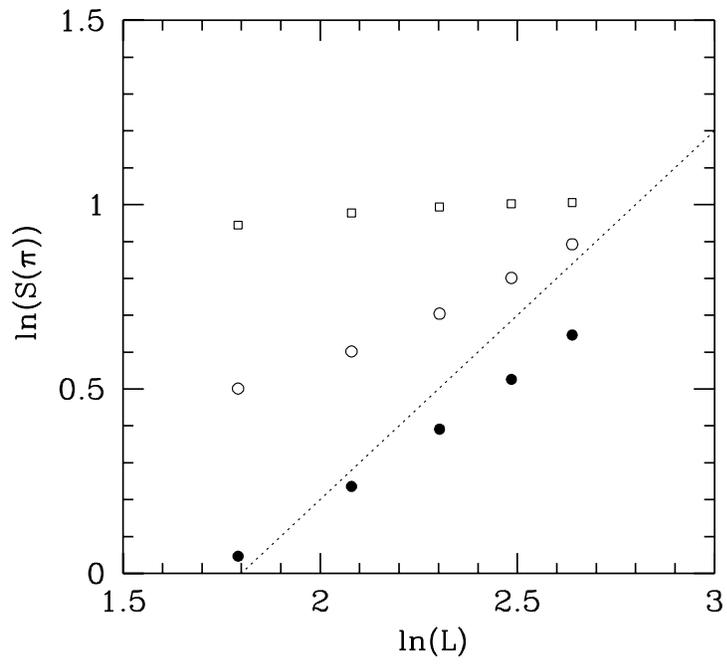
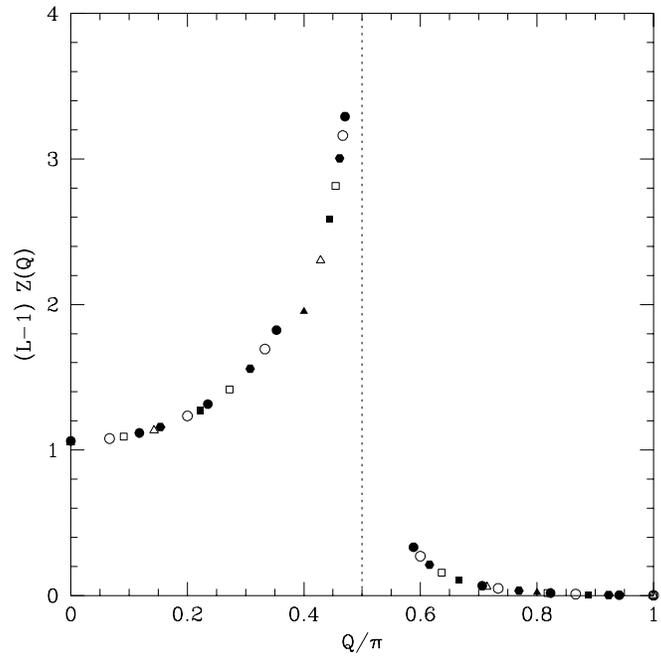
\end{document}